Creativity: Linchpin in the Quest for a Viable Theory of Cultural Evolution

Liane Gabora

University of British Columbia

For correspondence:

Liane Gabora
Department of Psychology, University of British Columbia
Okanagan Campus, Fipke Centre for Innovative Research
3247 University Way
Kelowna BC, Canada V1V 1V7
liane.gabora@ubc.ca

**Abstract**

This paper outlines the implications of neural-level accounts of insight, and models of the conceptual interactions that underlie creativity, for a theory of cultural evolution. Since elements of human culture exhibit cumulative, adaptive, open-ended change, it seems reasonable to view culture as an evolutionary process, one fueled by creativity. Associative memory models of creativity and mathematical models of how concepts combine and transform through interaction with a context, support a view of creativity that is incompatible with a Darwinian (selectionist) framework for cultural evolution, but compatible with a non-Darwinian (Self-Other Reorganization) framework. A theory of cultural evolution in which creativity is centre stage could provide the kind of integrative framework for the behavioral sciences that Darwin provided for the life sciences.

**Highlights**

- Other species create but we alone exhibit cumulative open-ended cultural evolution.
- A theory of cultural evolution could provide a unifying framework for the social sciences.
- Creative thought is incompatible with a Darwinian framework for cultural evolution.
- Creativity *is* compatible with evolution through self-organization and communal exchange.
- Cultural novelty is generated when concepts manifest anew in response to context.

**Introduction**

There is literature on cross-cultural differences in creativity [1–3], the adaptive value of creativity and how human creativity evolved [4–7], as well as efforts to frame creativity as a Darwinian [8,9][1] and a nonDarwinian [11,12] evolutionary process. However, with some exceptions [13–15], there is a dearth of research on the implications of how the creative process works for the question of *how culture evolves*. This appears to be an outstanding gap in the literature given that creativity is what *fuels* cultural evolution, and a theory of cultural evolution could provide an integrative framework for the social sciences in much the same way that fragmentary biological knowledge was unified by Darwin's theory of natural selection (and subsequently unified further by the neo-Darwinian synthesis, and research on epigenetic processes and complex systems [16,17]). This paper outlines how creativity research can contribute to this important enterprise.

**A new direction for creativity research**

Creative ideas are sometimes conceived of as discreet, separate entities much like objects in the physical world that can we search for and select amongst [8,9]. However, models of the contextual aspects of higher cognition [18–23], including concept combination and creativity [24–27], buttressed by neural-level accounts of memory and insight [28–30], point to a different view. This research suggests that thoughts and ideas are not separate and distinct but exist as part of an interwoven matrix until the instant you think of them. Moreover, each time you think of them they are reconstructed anew and you experience them differently, depending on the context, your recent experience, and what you have done and thought about since the last time you brought them to mind. Like Schrödinger's famous cat that is neither dead nor alive, a concept or unborn idea—when you're not thinking about it—neither exists nor does it not exist. It is in a what is called a *ground state*, a state of *potentiality*, and requires a context—something that *brings it to mind*—to *actualize* it. Much as if you shine light on an object from one direction it casts one shadow, and if you shine light from a different direction it casts another, the first time you try to articulate a creative idea it manifests as one output, and after thinking about it, it may manifest as a different output. Just because these two external realizations of the idea take different physical forms, that doesn't mean there are two discrete representations in the mind. They may be different realizations of the same underlying idea at different phases of a creative honing process.

Extending these ideas further lead to a new conception of the creative process. While the divergent and convergent phases of the creative process are often characterized respectively as *generative* and *evaluative* [31–33], associative memory models of creativity and mathematical models of how concepts combine and transform through interaction with a context suggest that phases of the creative process instead be characterized in terms of potentiality and actualization. In this view, the creative process begins with the recognition that one's understanding of something is in a state of potentiality—i.e., vague, ill-defined, or engendering emotional turbulence—so one examines it from different angles to better understand it. This *may* involve the emergence of new candidate ideas, but also it may not; it may simply entail a sharpening of the originally vague idea. Evaluation is occurring throughout; the entire process of reflecting on an idea consists of interactions between your current conception of it, and contexts you throw at it, and with each 'reflection' (interaction between idea and context) you evaluate the outcome. In the *divergent* phase of the creative process one reflects on the idea by considering it from *unconventional contexts*, and our ability to do this hangs on their ability to reform anew each time you think of them, as discussed above. In the latter *convergent* phase, the idea is refined by considering it in more *conventional contexts*, often generated through simulation of how others will receive it.

---

[1] However, previous supporters have backed away from this position, e.g., Simonton [12] has conceded that his theory's "explanatory value does not depend on any specious association with Darwin's theory of evolution".

While these views on creativity are nascent, as we will see following a brief examination of the workings of evolutionary processes and culture in particular, *they have implications for the question of how culture evolves*, and could play a vital role in the development of a viable theory of cultural evolution.

## Cultural evolution as a unifying framework for the social sciences

Darwin's theory of evolution by natural selection vastly enhanced our understanding of the organismal world by integrating scattered biological knowledge into a unified "tree of life". Since art, technology, languages, and customs change over time in a manner seemingly reminiscent of biological evolution, it seemed reasonable to view culture as a second evolutionary process, fueled in this case by human creativity. Although other species exhibit both creativity and imitation, humans build on each others' ideas, adapting them to our own tastes, needs, and desires, such that the process is open-ended, i.e., the space of possibilities cannot be pre-specified [34]. Thus, cumulative, adaptive, open-ended cultural evolution appears to be uniquely human.

There is a long history of attempts to frame cultural evolution as a Darwinian evolutionary process [35], and although highly contentious, the approach is still widespread [36–38]. A Darwinian process consists of two components: the *generation* of new variants, and the differential survival or *selection* of some of those variants, such that they live long enough to produce offspring. Since Darwin's explanation focused not on the generation of variants but on the selection of some fraction of them, it can be referred to as a *selectionist* theory. Darwin posited that biological change is due to the effect of differential selection on the distribution of randomly generated heritable variation in a population over generations; in other words, 'survival of the fittest'. Organisms with adaptive traits have more offspring—i.e., are 'selected' for—and therefore, their traits proliferate over time. Notice that the theory operates on the timescale of generations, as it requires at a minimum a generation for change to occur. Note also that it assumes that variants are separate and distinct entities that can be selected amongst such that some survive and others do not.

Dawkins [39] proposed that natural selection requires a *replicator*, which he defined as an entity with the following characteristics: *fecundity* (it replicates), *longevity* (it survives long enough to replicate), and *fidelity* (after several generations of replication, it is still almost identical to the original). Holland [40], a pioneer in the field of complex, adaptive systems, showed that this is necessary but not sufficient for a selectionist explanation to hold, and provided a more nuanced analysis of what is required (Table 1).

|  | **Replicator (Dawkins)** | **Self-assembly Code (Holland)** |
|---|---|---|
| Self-replication | Yes | Yes |
| Fecundity; longevity; fidelity | Yes | Yes |
| Passive copying and active transcription of self-assembly instructions | ? | Yes |
| Sequestration of inherited information | ? | Yes |
| Genotype / phenotype distinction | ? | Yes |
| Transmission of acquired traits | ? | No |
| Evolutionary processes it seeks to explain | Biological, cultural | Biological |

Table 1. Comparison of Dawkin's view that natural selection requires a replicator versus Holland's view that it requires a self-assembly code. Both involve self-replication with fecundity, longevity, and fidelity. Only the self-assembly code requires instructions for generating a copy of the self that are both passively copied and actively transcribed. As a result, only this view is committed with respect to (1) the sequestration of inherited information, (2) a clear distinction between genotype and phenotype, and (3) a prohibition on transmission of acquired traits. The replicator has been proposed as the central construct of an evolutionary framework for both biological and cultural evolution. Since cultural evolution lacks self-assembly instructions that are both passively copied and actively transcribed, the self-assembly code can function as the central construct for biological evolution only.

Holland's first requirement is *randomly generated variation*. A selectionist process works through competitive exclusion amongst *existing* variants; it does not work by affecting how variants are *generated*. In fact, to the extent that variation is not generated randomly, the distribution of variants reflects *whatever was biasing the generation away from random in the first place*, rather than survival of the fittest.

Holland's second requirement is *negligible transmission of acquired traits*. The reason a selectionist theory is applicable in biology is that there are two kinds of traits: (1) *inherited traits* (e.g., blood type or eye color), which are transmitted *vertically* from parent to offspring by way of the genes, and (2) *acquired traits* (e.g., a tattoo, or knowledge about pop stars), which are obtained during an organism's lifetime, and which are sometimes called *epigenetic* because they are transmitted *horizontally* from outside sources, not vertically by way of genes. Because acquired traits are not passed down (e.g., you do not inherit your mother's tattoo), the fast, *intra*-generational transmission of acquired traits does not drown out the slow, *inter*-generational transmission of inherited traits. In other words, a selectionist explanation works when acquired change is negligible relative to inherited change; otherwise the first, which can operate in an instant, overwhelms the second, which requires generations. Thus, replicators could *exist* and *evolve*—i.e., exhibit cumulative, adaptive, open-ended change—but to the extent that variation was generated non-randomly, or that there was non-negligible transmission of acquired traits, their evolution could not be accounted for by the theory of natural selection; another theory is required to explain how they evolve.

We know of no means of avoiding the transmission of acquired traits other than by way of a *self-assembly code* (such as the genetic code), i.e., a set of instructions for how to reproduce. Thus, the third requirement is *precisely-orchestrated expression of a self-assembly code*. In the case of biological evolution this code takes the form of nucleotide pairs that make up our DNA[2]. In theory, it could take another form.[3] But whatever this code is made of, *its low-level information-bearing components must be organized in an orderly, predictable fashion* so as to be amenable to *parsing into meaningful units* and thereby *avoid disrupting the precisely orchestrated process by which it is expressed to generate offspring*. In the case of biology these units are genes, which (by way of elaborate transcription and translation processes) are interpreted to make offspring. The genetic information in parental DNA must be similar enough at the lowest level of the basic building blocks of which they are composed (i.e., nucleotide pairs), otherwise they will not be parsed properly into genes when they are interpreted to make the traits, organs, systems, and so forth that comprise a viable offspring.

      Before examining whether cultural evolution meets Holland's requirements for a selectionist explanation, we note that they are not always met in biology [16], though for many biological processes they are met sufficiently well that Darwin's theory serves as an adequate explanatory model. Culture, however, is another matter [41]. First, *cultural variation is not randomly generated*. Darwinian cultural theorists sometimes concede this point [38], but fail to see that (1) this invalidates a Darwinian theory of culture (as explained above), and (2) a non-Darwinian evolutionary theory of culture is possible (as explained below). Second, *in cultural evolution, acquired traits are transmitted*. Unlike biological evolution, in cultural evolution there is no mechanism by which changes acquired over a lifetime are shed on a regular basis at the end of each generation (e.g., once one cup had a handle, all cups could have handles). Therefore, acquired (horizontal) change can accumulate orders of magnitude faster than, and overwhelm, vertical change due to the mechanism Darwin proposed: differential replication of heritable variation in response to selection over generations. Third, *culture lacks the precisely-orchestrated expression of a self-assembly code*. It is not the cultural artifact itself that spontaneously reproduces; it is human minds that make that happen. Even an artifact such as a blueprint that consists of coded assembly instructions is not a *self*-assembly code, i.e., it does not spontaneously reproduce new offspring blueprints. Indeed if, as discussed above, variants are not separate and distinct amongst such that some survive and others do not, but rather reassembled anew each time they are brought to mind, the replication of ideas can be highly *imprecise*. A thought or idea can merge with one or more other thoughts or ideas to produce something altogether new. For example, a playground designer might reconceive of a tire as a swing seat while a landscaper might pile old tires to make a retaining wall (Figure 1). This kind of extreme imprecision would not be viable in a system that relies upon low-level information-bearing components being organized in an orderly, predictable fashion so as to be parsed into meaningful units and expressed to generate offspring.

---

[2] Although Darwin did not know of the existence of DNA, he was aware that *some* underlying mechanism was resulting in a prohibition on transmission of acquire traits.

[3] As per the substrate neutrality argument.

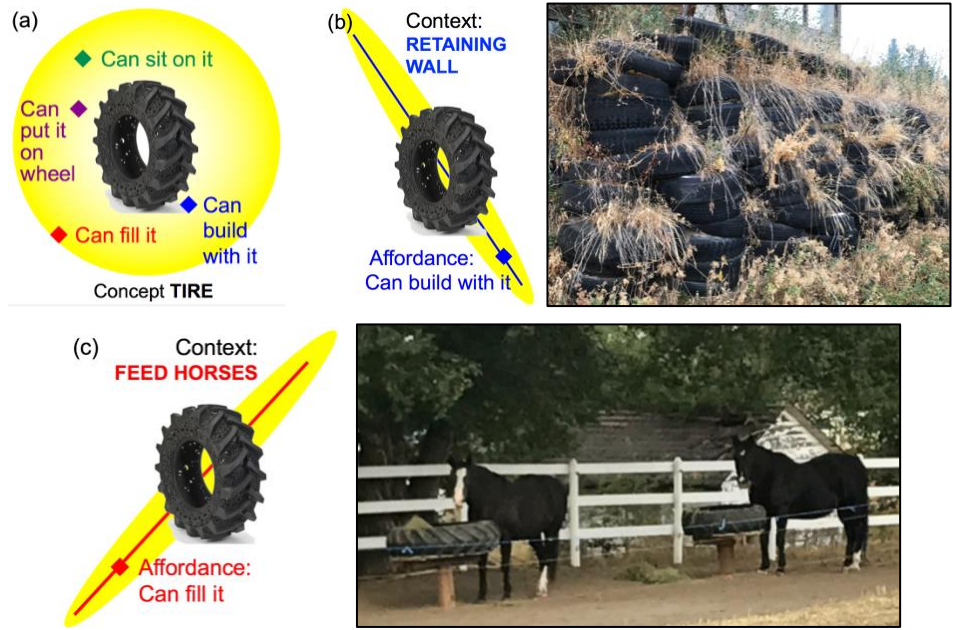

Figure 1. Cultural evolution of what to do with a tire. The concept TIRE can be adapted to these different contexts because when it is not thought about it exists in a 'ground state' of potentiality (a), and reforms anew when brought to bear on a particular situation (b and c). For a neighbor who needed a retaining wall, the affordance 'can build with it' enabled the concept TIRE to be reconceived of as a brick, while for a neighbour with horses the affordance 'can fill it' enables it to be thought of as a food bowl. (Photos taken by author.)

This analysis precludes a *selectionist* but not an *evolutionary* framework for culture. The vital importance of non-Darwinian evolutionary processes is increasingly recognized [16,17,42]. Research on the origin of life suggests that early life consisted of autocatalytic protocells that evolved through a non-Darwinian process, and natural selection emerged later from this more haphazard ancestral evolutionary process [43–46]. This non-Darwinian process requires (1) a *self-organizing network* of components that generate new components through their interactions, (2) the network should be able to reconstitute another like itself through haphazard (not code-driven) duplication of components, and (3) interaction amongst networks. This process can be referred to as *Self-Other Reorganization* (SOR) because it involves an interplay between self-organized *internal* restructuring and communal exchange *amongst* autocatalytic structures. Like Darwinian evolution it has mechanisms for preserving continuity and for introducing novelty, but unlike Darwinian evolution it is a low-fidelity Lamarckian process (Table 2).

|  | **Natural Selection** | **Self-other Re-organization (SOR)** |
|---|---|---|
| Unit of self-replication | DNA | Self-organizing network |
| Mechanism for preserving continuity | Reproduction (vertical transmission), proofreading enzymes, etc. | Communal exchange (a form of horizontal transmission) |
| Mechanism for generating novelty | Mutation, recombination | *Creativity and innovation,* transmission error |
| Self-assembly code | Yes | No |
| High fidelity | Yes | No |
| Transmission of acquired traits | No | Yes |
| Type of process | Darwinian | Lamarckian (by some standards) |
| Evolutionary processes it can explain | Biological | Early life; horizontal gene transfer (HGT), cultural |

Table 2. Similarities and differences between two evolutionary frameworks: natural selection and SOR. Both have mechanisms for preserving continuity and introducing novelty. However, whereas natural selection is a high fidelity Darwinian process and the structure that self-replicates is self-assembly instructions, communal exchange is a low fidelity Lamarckian process, and the structure that replicates is a self-organizing network. Only SOR allows transmission of acquired traits. SOR is proposed to be the mechanism by which early life evolved as well as the mechanism by which culture evolves, and some aspects of present day life, such as horizontal gene transfer.

It has been proposed that, as did early life, culture evolves through SOR [47–53; see 54 for a related approach]. Here, the self-organizing networks are not protocells exchanging catalytic molecules, but minds exchanging ideas. As parents and others share knowledge with children, an integrated understanding of the world takes shape in their minds, and they become creative contributors to cultural evolution.

I emphasize that it is not claimed that Darwin's theory of natural selection is incorrect, nor that it does not provide an adequate explanatory framework for biological evolution, merely that it does not provide an adequate explanatory framework for cultural evolution.[4]

## Challenges from creativity research for a viable theory of cultural evolution

Despite calls to identify the cognitive mechanisms underlying cultural evolution [56,57], absent amongst the eight challenges posed in a recent effort to establish 'grand challenges' for cultural evolution research [57] is the challeng of *understanding the creative processes that drive cultural evolution*. However, creativity research is well poised to play a prominent role in establishing how culture evolves by offering a litmus test for a viable theory of cultural evolution. Let us propose the following challenges: a viable theory of cultural evolution must be able to explain and describe (1) Banksy's expression of political convictions through art [59], (2)

---

[4] Holland himself once used the term "Darwinian selection" loosely to refer to a situation in which "agents that collect resources more rapidly than others contribute more of their characteristics to future generations" [55, p. 56]. However, in this passage he is not trying to provide a comprehensive definition of what Darwinian selection entails as he was in [40]. He may well not have been aware of the extensive work by Carl Woese and others showing that it is possible to evolve through means other than Darwinian selection [43–46]. (Unfortunately since he is no longer with us we cannot ask him to clarify this.)

Led Zeppelin's use of *Lord of the Rings* by Tolkien as inspiration for the songs "Battle of Evermore" and "Ramble On" [60], and (3) the 'spinning' of a news story to make it consistent with political convictions [61,62].

A selectionist framework cannot begin to address such challenges because it requires that variation be randomly generated; it cannot account for the generation of ideas through strategy or intuition or thinking something through for oneself. Thus, it cannot account for processes in which an idea *acquires* non-random change *within* a mind (e.g., the idea behind *Lord of the Rings* transforming from a book to a song) between events in which communicated *between* minds (i.e., between reading the book and re-expressing it as a song). Since it requires a predictably organized self-assembly code so as not to disrupt the precisely orchestrated process of reproduction, it cannot accommodate findings such as that cross-domain inspiration is ubiquitous in creative thought [63]. (If a melody inspires a painting, for example, there is no self-assembly code in this 'lineage' driving a precisely orchestrated reproduction process.) Many studies that claim to be about cultural *evolution*, such as studies of conformity bias [64], actually concern *transmission*. While transmission—the spread of *existing* traits—is a component of evolution, evolution additionally entails cumulative, adaptive, open-ended change. Studies that do incorporate cultural novelty are often limited to trivial sources such as 'cultural mutation' [65] or copying error [66], thereby neatly avoiding the issue of creativity.

SOR *can* accommodate these challenges, because it does not require that novelty be randomly generated, and it lacks the key signature of a Darwinian process: prohibition on transmission of acquired traits. It is also consistent with the new view of creativity discussed earlier. Recall the notion that an unborn idea exists in a ground state of potentiality and requires a context to bring it to mind or actualize it. For any concept or idea, there exists *some* context— some perspective or lens from which you could view it—that would motivate combining it with any other concept or idea. In other words, because our minds are integrated networks there is some possible situation that would inspire you to merge, say, the concepts 'duck' and 'piano'. Since no combinations are a priori off limits we start to see how human creativity could be open-ended. Moreover, if ideas are not discrete and separate until actualized by a context, it makes sense that the mind is what evolves through culture.

## Conclusions

This article provided an overview of the rationale for, and current status of, the application of creativity research to the question of how culture evolves. The open-ended, cumulative nature of human creativity presents formidable challenges for a theory of cultural evolution. A Darwinian (selectionist) explanation requires that transmission of *acquired* traits be negligible so that change over generations are explicable in terms of selection of randomly generated heritable variation. We know of no means of accomplishing this other than by way of a self-assembly code, which requires low-level predictability of information-bearing components to ensure precise orchestration of the intricate processes culminating in reproduction. Culture does not meet these requirements because human creativity is non-random, ideas acquire change as people mull them over, and there is no self-assembly code. SOR, a lesser-known evolutionary framework for cultural evolution which does not require randomly generated novelty and allows transmission of acquired traits, appears consistent with current directions in creativity research.

In sum, as those most knowledgeable about the creative processes at the core of cultural evolution, no group of scholars is better positioned than creativity researchers to unravel the mystery of how culture evolves.

**Funding**

The author acknowledge funding from grant (62R06523) from the Natural Sciences and Engineering Research Council of Canada.

**Conflict of interest statement**

Nothing declared.

**Acknowledgements**

The author thanks Victoria Scotney for assistance with the manuscript.

**References and recommended reading**

Papers of particular interest, published within the period of review, have been highlighted as

● of special interest

●● of outstanding interest